\documentclass[fleqn,usenatbib]{mnras}

\usepackage{newtxtext,newtxmath}
\usepackage[T1]{fontenc}
\usepackage{ae,aecompl}

\usepackage{bm}
\usepackage{graphicx}	% Including figure files
\usepackage{amsmath}	% Advanced maths commands
\title[Hydrodynamic instability in a warped disc]{Parametric instability in a free evolving warped protoplanetary disc}

\author[Deng, Ogilvie \& Mayer]{
Hongping Deng, $^{1}$\thanks{E-mail:hd353@cam.ac.uk}
Gordon I. Ogilvie, $^{1}$\thanks{gio10@cam.ac.uk} Lucio Mayer $^{2}$ \\
$^{1}$ Department of Applied Mathematics and Theoretical Physics, University of Cambridge, Centre for Mathematical Sciences,\\
Wilberforce Road, Cambridge CB3 0WA, UK\\
$^{2}$Center for Theoretical Astrophysics and Cosmology, Institute for Computational Science, University of Zurich,\\
Winterthurerstrasse 190, 8057 Zurich, Switzerland
}

\pubyear{2018}

\begin{document}
\label{firstpage}
\pagerange{\pageref{firstpage}--\pageref{lastpage}}
\maketitle

\begin{abstract}

Warped accretion discs of low viscosity are prone to hydrodynamic instability due to parametric resonance of inertial waves as confirmed by local simulations. Global simulations of warped discs, using either smoothed particle hydrodynamics (SPH) or grid-based codes, are ubiquitous but no such instability has been seen. Here we utilize a hybrid Godunov-type Lagrangian method to study parametric instability in global simulations of warped Keplerian discs at unprecedentedly high resolution (up to 120 million particles). In the global simulations, the propagation of the warp is well described by the linear bending-wave equations before the instability sets in. The ensuing turbulence, captured for the first time in a global simulation, damps relative orbital inclinations and leads to a decrease in the angular momentum deficit. As a result, the warp undergoes significant damping within one bending-wave crossing time. Observed protoplanetary disc warps are likely maintained by companions or aftermath of disc breaking.
\end{abstract}


\section{Introduction}
Warped accretion discs are common around black holes and stars. Discs may naturally form warped owing to the accretion of gas with different angular momenta \citep[see, e.g.,][]{Lucas2013,Bate2018}.  Initially planar discs can be warped by external torques, for example, the Lense--Thirring torque from a misaligned spinning black hole at the centre of the disc \citep{Bardeen1975} or the gravitational torque due to the quadrupole moment of the central body \citep{Tremaine2014} or misaligned companions \citep{PapaloizouTerquem1995,Xiang-Gruess2013}. Magnetic fields \citep{Lai1999} and radiation pressure \citep{Pringle1996} may also cause disc warps. 

A warped disc was first inferred in the binary X-ray source Hercules X-1 \citep{Katz1973,Gerend1976}, and the superorbital modulation of numerous X-ray binaries is now attributed to precessing warped discs \citep{Kotze2012}. Warps have been observed in maser emission from accretion discs around supermassive black holes in the centres of active galaxies such as NGC~4258 \citep{Miyoshi1995}. There is growing indirect observational evidence for warped circumstellar discs around young stars, indicated by shadows cast by the warps \citep{Rosenfeld2012,Marino2015,Casassus2018,Stolker2016,Benisty2018}. A moderate warp has also been directly detected in the young circumstellar disc around the protostar IRAS 04368+2557 \citep{Sakai2019}. The growth of well-resolved observations of warped discs necessitates a comprehensive theory of these objects.

The quantitative study of warped discs subject to external torques began with \cite{Bardeen1975}. \cite{Petterson1977} and others derived partial differential equations governing the evolution of twisted fluid rings. However, these early attempts were flawed, especially because they considered only the action of viscous stresses and neglected the internal flows driven by pressure gradients in a warped disc. \cite{Papaloizou1983} realized the importance of angular momentum advection by the oscillatory horizontal flow whose amplitude is determined by the viscosity. They derived the equations for the rapid diffusion (at a rate inversely proportional to the viscosity) of a small-amplitude warp in a Keplerian disc when the dimensionless viscosity parameter $\alpha$ \citep{Shakura1973} is larger than the disc's aspect ratio $H/R$ ($H$ being the scale-height at radius $R$). In a low-viscosity Keplerian disc, the warp behaves instead as a bending wave travelling at a speed related to the sound speed \citep{Papaloizou1995}. Based on angular momentum conservation, \cite{Pringle1992} formulated a simplified non-linear framework governing the the evolution of the surface density $\Sigma$ and the local orbital tilt vector $\bm{l}(r,t)$ \citep[see also][]{Papaloizou1983}. \cite{Ogilvie1999} derived a fully non-linear theory of warped discs in the diffusive regime, presenting general formulae for the torques associated with the internal flows. Two of the torque components are encapsulated by the simple viscosity parameters in \citet{Pringle1992}, although the coefficients are not constant, while the third one causes the warp to precess or twist.

The oscillatory horizontal flow in a warped disc can be unstable as a result of parametric resonance of inertial waves, as anticipated by \cite{PapaloizouTerquem1995} and first shown in detail for Keplerian inviscid discs by \cite{Gammie2000}, using a local shearing-box model in which the oscillatory flow was initiated and allowed to decay. \cite{Ogilvie2013a} developed a warped shearing box model for slowly evolving warps, in which the horizontal flow is forced periodically by the local geometry of the warped disc. The parametric instability was found to be widespread for warped discs \citep{Ogilvie2013b} in the local warped shearing box model, as was further confirmed in numerical simulations by \cite{Paardekooper2019}, although regimes were found in which the internal shear flows are so strong that no linear instability can be clearly identified. For completeness, we also note that strong magnetic fields can detune the hydrodynamic resonance between the epicyclic and vertical oscillation frequencies, and that more complicated resonances may then occur \citep{Paris2018}.

The dynamic boundary of a warped disc poses a challenge for grid-based codes. Many previous warped disc simulations employed Smoothed Particle Hydrodynamics (SPH) \citep{Lucy1977,Gingold1977}. As a Lagrangian method, SPH tracks the evolution of fluid elements and naturally adapts to the complex flow geometry. \cite{Nelson1999} verified the existence of the distinct diffusive and bending wave regimes with global SPH simulations. \cite{Lodato2010} carried out a suite of high-resolution SPH simulations (up to 20 M particles) and found good agreement between the simulation and the non-linear theory of \cite{Ogilvie1999} in the diffusive regime.   Many other simulations have included more complex physics, such as companions \citep{Xiang-Gruess2013,Facchini2013, Fragner2010} and Lense--Thirring torques \citep{Nixon2012,Nealon2015}. However, no simulation has reported turbulence induced by the parametric instability except the dedicated high-resolution local simulations of \citet{Gammie2000} and \citet{Paardekooper2019}.

We may wonder whether the parametric instability occurs in a global,  evolving warp or exists only in an idealised local model with periodic radial boundary conditions. If the warp is localized in radius, or otherwise spatially inhomogeneous, then the global interaction of the travelling inertial waves with the warp may need to be considered. It is also possible that the low resolution and high numerical viscosity of previous global simulations have suppressed the instability. The grid-based simulations by \cite{Fragner2010} and \cite{Sorathia2013} used fewer than 10 cells per scale-height compared to 32 or more cells per scale-height in \cite{Paardekooper2019}. SPH employs artificial viscosity to capture shocks and prevent particle disorder \citep[see, e.g., the review by][]{Price2012}. Artificial viscosity can be wrongly triggered and lead to excessive dissipation \citep{Cullen2010,Bauer2012}; many improvements have been proposed recently \citep[e.g.][]{Rosswog2015} but were not used in previous SPH simulations of warped discs.

New hybrid hydrodynamic methods are thriving in computational astrophysics, e.g.\ moving mesh methods \citep{Springel2010} and Godunov-type Lagrangian methods \citep{Hopkins2015}. Among the latter,  the Meshless Finite Mass (MFM) method has been shown to be less viscous than SPH with little extra computational cost. MFM captures shocks without introducing artificial viscosity and thus shows better angular momentum conservation than SPH \citep{Hopkins2015,Deng2017}. At sufficiently high resolution it can even capture turbulence resulting from the magnetorotational instability while SPH fails and leads to unphysical growth of the magnetic field \citep{Deng2019,Deng2020}. MFM appears promising for modelling low-viscosity warped discs.

We apply this novel Godunov-type Lagrangian method to simulate warped discs, focusing on the development of the parametric instability. The remainder of the paper is organised as follows. In Section~\ref{sec:model} we recall the bending-wave theory and the disc model we intend to study. We detail how we realize the disc model with our code in Section~\ref{sec:code}. The disc evolution and evidence for the parametric instability are presented in Section~\ref{sec:results}. We discuss the astrophysical implications in Section \ref{sec:dis} and conclude in Section~\ref{sec:con}.

\section{Disc model}
\label{sec:model}

Ideally we would set up a steady warped disc and look for growing modes of the parametric instability \citep{Gammie2000,Ogilvie2013b}. \citet{Tremaine2014} provided a series of steady-state solutions to the equations for viscous warped discs subject to quadrupolar or Lense--Thirring torques from the central body and a tidal torque from an outer companion. Technically, these solutions are difficult to realise with a Lagrangian code, and the low-viscosity solutions in which we are most interested are especially hard to represent with our hydrodynamic code. 
Physically, we are more interested in isolated discs around single stars, and no steady warped state is possible in the absence of external torques. We therefore choose to set up a freely evolving warp similar to \cite{Lodato2010}. Whether parametric instability happens in an evolving warp and how it affects the warp evolution have not been explored. Previous local simulations rely on the assumption that the warp evolution is much slower than the development of the instability \citep{Gammie2000,Paardekooper2019}.

The problem we study is essentially scale-free. However, we describe the solutions in physical units with an application to protoplanetary discs in mind. We study a disc of 0.05 solar mass ($0.05\,M_\odot$, although the self-gravity of the disc is ignored) around a solar-mass star ($M=M_\odot$) with a surface density inversely proportional to the radius ($\Sigma\propto R^{-1}$). The disc extends from 5--50 au. The disk is vertically isothermal with a constant aspect ratio $H/R=0.05$ ($H$ being the scale-height at radius $R$). The evolution of a linear bending wave in an inviscid, Keplerian disc with angular velocity $\Omega=(GM/R^3)^{1/2}$ is governed by the following equations \citep[e.g.][]{Ogilvie1999,Lubow2000}:
%\begin{align}
%	&\Sigma R^2 \Omega\frac{\partial \bm{l}}{\partial t}=\frac{1}{R}\frac{\partial \bm{G}}{\partial R} +\bm{T} \\
%   & \frac{\partial\bm{G}}{\partial t}-\left(\frac{\Omega^2-\kappa^2}{2\Omega}\right )\bm{l}\times %\bm{G}+\alpha\Omega\bm{G}=\frac{PR^3\Omega}{4}\frac{\partial \bm{l}}{\partial R} 
%\end{align}
\begin{align}
   &\Sigma R^2 \Omega\frac{\partial \bm{l}}{\partial t}=\frac{1}{R}\frac{\partial \bm{G}}{\partial R}, \label{bending1} \\
   &\frac{\partial\bm{G}}{\partial t}=\frac{PR^3\Omega}{4}\frac{\partial \bm{l}}{\partial R}, \label{bending2}
\end{align}
where $\bm{l}(R,t)$ and $\bm{G}(R,t)$ are the horizontal components of the unit tilt vector and the internal torque, respectively, and $P(R)$ is the vertically integrated pressure, related to the isothermal sound speed $c_s(R)$ and scale-height $H(R)$ by $(P/\Sigma)^{1/2}=c_s=H\Omega$. These equations can also be written in a complex form, using the tilt and torque variables $W=l_x+il_y$ and $\mathcal{G}=G_x+iG_y$ \citep[cf.][]{Pringle1996}, which are further related to the radial velocity and enthalpy perturbations, $u^\prime_R=\mathrm{Re}[Az\,e^{i\phi}]$ and $w^\prime=\mathrm{Re}[Dz\,e^{i\phi}]$, through $W=-D^*/R\Omega^2$ and $\mathcal{G}=\frac{1}{2}\Sigma H^2R^2\Omega A^{*}$. We solve the equations
%\begin{align}
%&\frac{\partial A}{\partial t}=-\frac{D}{R}-\frac{1}{2}\frac{\partial D}{\partial R} + i\eta \Omega A -\alpha %\Omega A \\
%&\frac{\partial D}{\partial t}=-\frac{c_s^2}{2}\left[\frac{1}{\Sigma H^2 R^{1/2}}\frac{\partial}{\partial %R}(\Sigma H^2 R^{1/2}A)\right] +i\zeta \Omega D,
%\end{align}
\begin{align}
&\frac{\partial A}{\partial t}=-\frac{D}{R}-\frac{1}{2}\frac{\partial D}{\partial R},  \\
&\frac{\partial D}{\partial t}=-\frac{c_s^2}{2}\left[\frac{1}{\Sigma H^2 R^{1/2}}\frac{\partial}{\partial R}(\Sigma H^2 R^{1/2}A)\right]
\end{align}
for $A$ and $D$ following the method of \cite{Lubow2002}. 

\begin{figure}
\centering
\includegraphics[width=\linewidth]{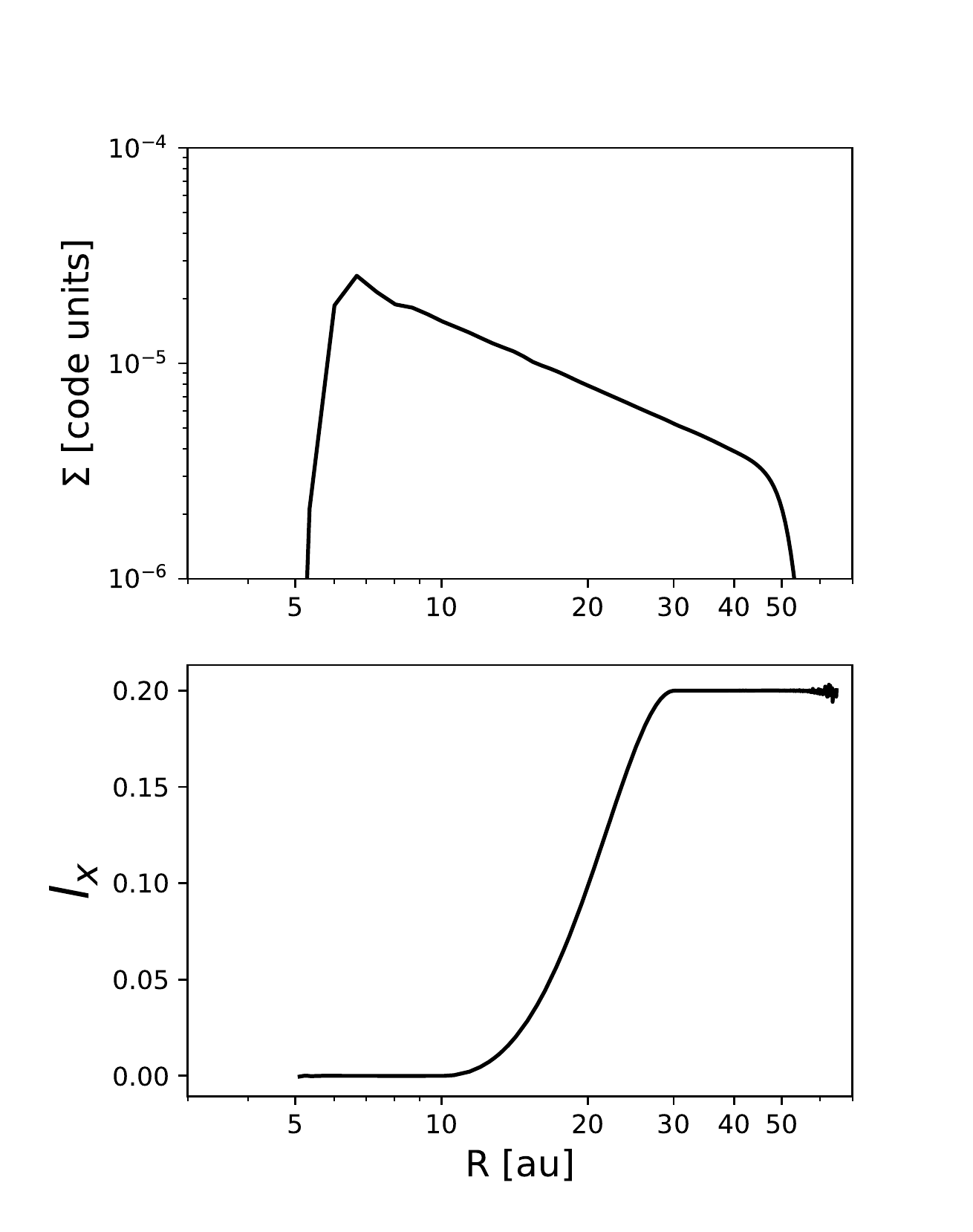}
\caption{The initial surface density profile and $l_x$ profile. There is a diffusive tail extending to 60~au.  \label{fig:ic}}
\end{figure}

Here we initialize an untwisted warp, i.e. $l_y=0$ and
\begin{equation*}
l_x=
\begin{cases}
   0 ,                       & R \leq 10,\\
    0.1 \left[1+\sin \left(\frac{(R-20)\pi}{20}\right )\right],              & 10 < R <30, \\
    0.2,              & R \geq30.
\end{cases}
\end{equation*}
The initial condition is depicted in Fig.~\ref{fig:ic}. Note that we apply free boundary conditions at both radial edges of the disc and thus diffusive tails form naturally (see discussion below). 

Equations~(\ref{bending1})--(\ref{bending2}) can in fact be solved analytically for our power-law disc model by transforming them into the classical wave equation. The solution, derived in Appendix~\ref{sec:A}, shows how the initial condition resolves into inwardly and outwardly propagating bending waves that reflect repeatedly from the inner and outer boundaries. Our numerical bending wave solution agrees extremely well with this analytical solution. 

\begin{figure*}
\centering
\includegraphics[width=0.9\linewidth]{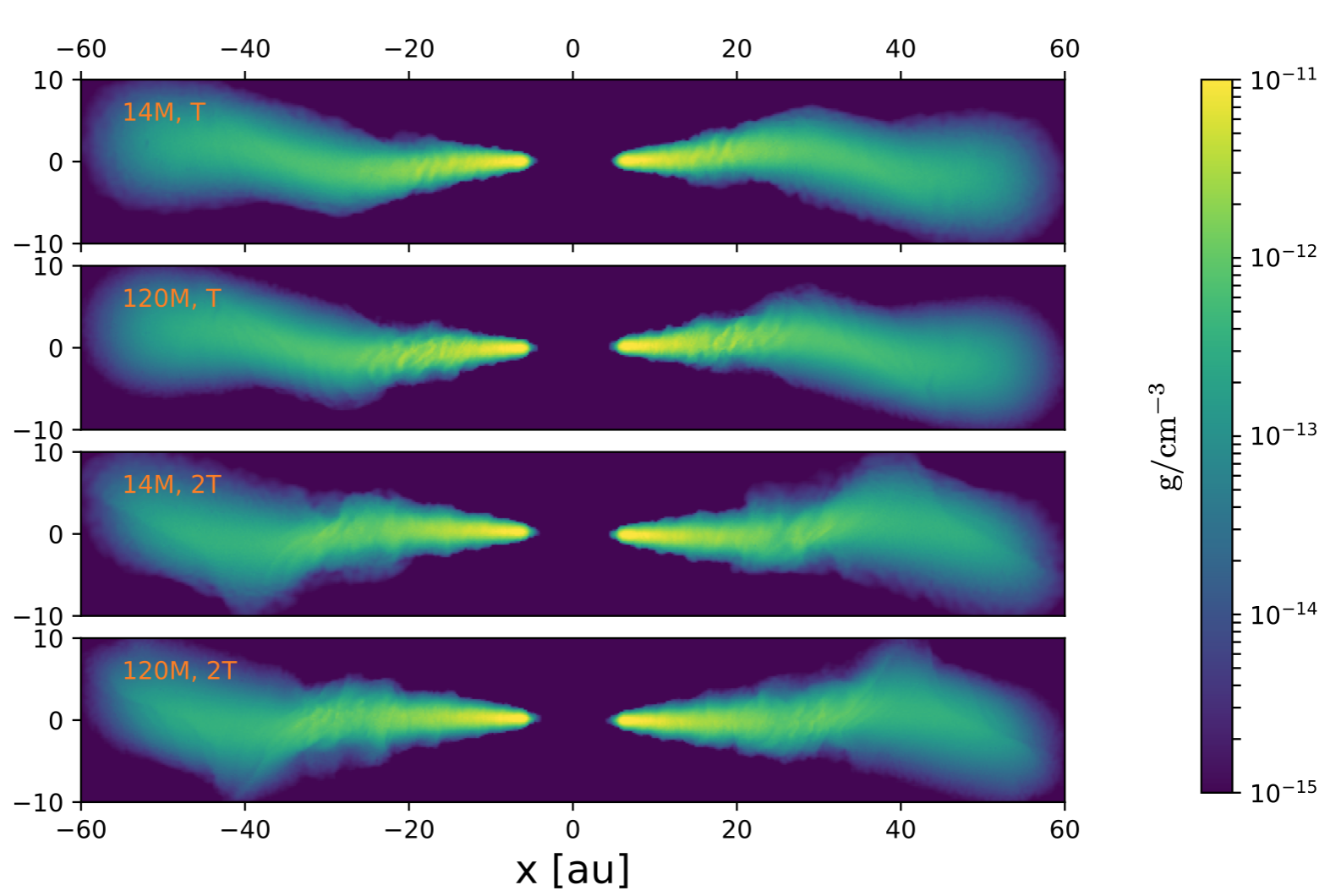}
\caption{Side-on volume density slice plots (at $y=0$) with labels indicating the total number of particles (14~M or 120~M) in the simulation and the snapshot time ($T$ equals 430 years). Here the vertical ($z$) axis is aligned with the disc's total angular momentum vector.\label{fig:rho}}
\end{figure*}

\section{Numerical method}
\label{sec:code}
\begin{figure}
\centering
\includegraphics[width=0.8\linewidth]{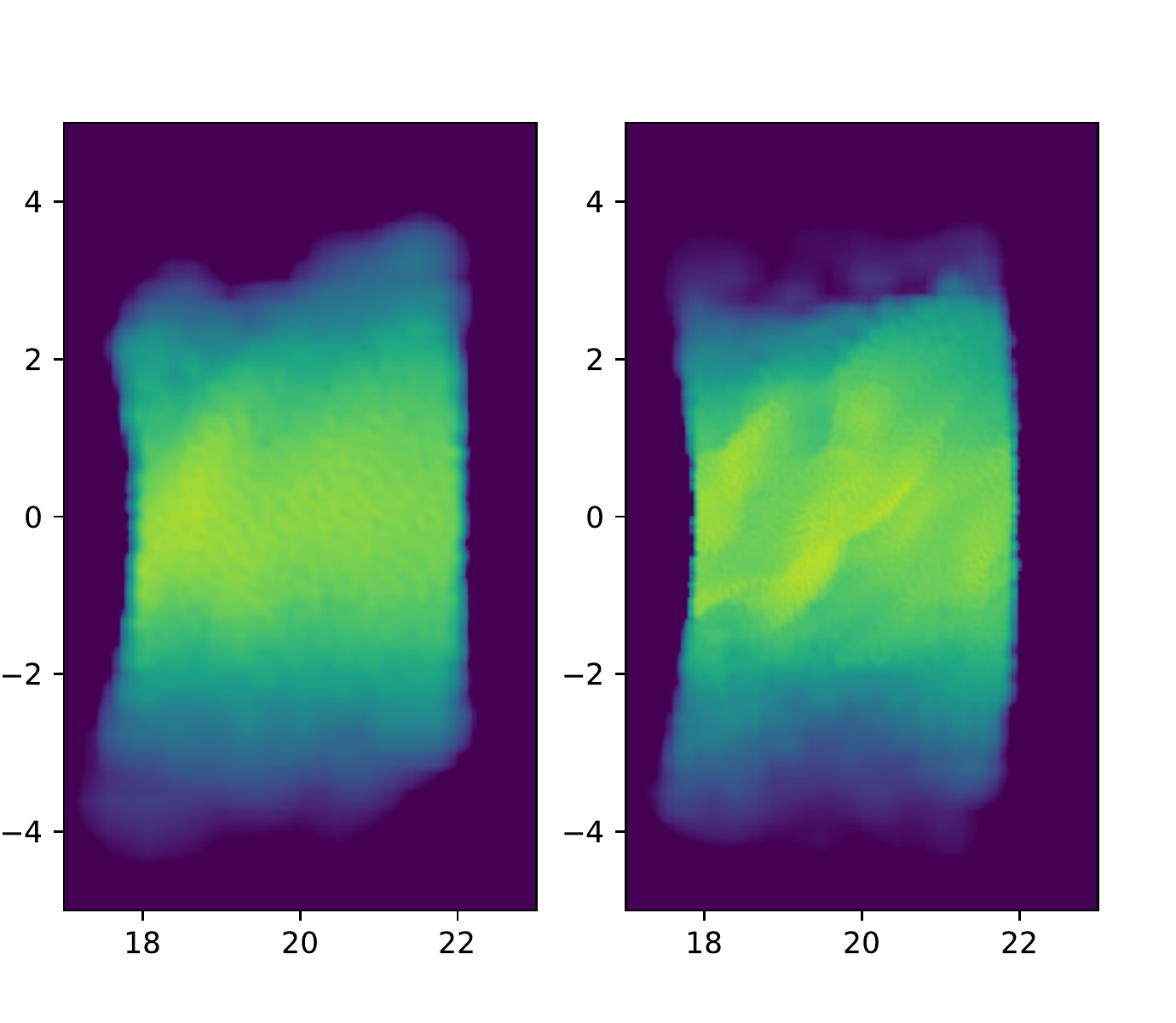}
\caption{Zoomed-in plots of the upper two panels of Fig.~\ref{fig:rho} (left: 14~M, right: 120~M). Here the vertical ($z$) axis is parallel to the local angular momentum vector.\label{fig:rholocal}}
\end{figure}

The parametric instability is sensitive to resolution and viscosity and has only been seen so far in local simulations \citep{Gammie2000, Paardekooper2019}. We used the Godunov-type Lagrangian method in the GIZMO code in the Meshless Finite Mass (MFM) mode \citep{Hopkins2015}. It does not employ the artificial viscosity used in previous warped disc simulations with Smoothed Particle Hydrodynamics (SPH) and thus shows better conservation properties than SPH \citep{Hopkins2015, Deng2017}. As mentioned in the Introduction, at sufficiently high resolution, MFM can even capture the subsonic turbulence resulting from the magnetorotational instability \citep{Deng2019,Deng2020}.

A resolution study is necessary before applying MFM for the first time to the study of warped discs. In Appendix~\ref{sec:B} we estimate the numerical viscosity by studying the decay of a long-wavelength bending wave in the local shearing-box model. The numerical $\alpha$ parameter \citep{Shakura1973} is shown to be below 0.001 at a resolution scale of $H/8$ in the disc midplane; note that we always adopt the Wendland C4 smoothing kernel to minimize numerical noise \citep{Dehnen2012, Deng2019}. At this resolution we can locally resolve the growth of the parametric instability similarly well as in the grid-code \citep[ZEUS; see][]{Stone1992} simulations with 32 cells per $H$ used by \cite{Gammie2000} (see Appendix~\ref{sec:C}). This resolution has never been achieved in previous global simulations of warped discs.

To save computational resources we utilized the particle splitting scheme in the GIZMO code to prepare initial conditions \citep[see also][]{Deng2020}.  We sample the density (Fig.~\ref{fig:ic}) profile by a Monte Carlo placement method \citep{Deng2017}. Initially, we build a planar disc model with 2~M particles and relax it to a steady state by gradually damping radial particle motion. The central star is treated as a sink particle that accretes any other particle within 5~au. The relaxation was run for 4 outer rotation periods (ORPs) and we verified that the relaxed disc remained steady for a further 4~ORPs, showing no signs of evolution.   The disc's surface density beyond 7~au is barely affected by the inner boundary. A low-density diffusive tail forms beyond 45~au and the density gradually approaches zero (Fig.~\ref{fig:ic}). We caution that the diffusive tail does not have Keplerian rotation. We then split the particles several times to reach the desired resolution, either 14~M or 120~M particles. We relax the two high-resolution planar discs for a short time to remove noise introduced by particle splitting. Eventually, we managed to keep the energy associated with particle noise about 10000 times smaller than the thermal energy; see Fig.~\ref{fig:Ekz}. Finally, we apply a radially dependent rotation matrix to the planar disc to realize the setup discussed in Section~\ref{sec:model}. 

We achieve a resolution of $H/8$ in the midplane in the 120~M particle simulation; note the disc's aspect ratio is constant so that the disc is equally resolved at all radii.  This simulation was run for one full bending wave crossing time ($\tau(50)-\tau(5)$ in equation \ref{eq:tau}). It took 600~K CPU hours on the Piz Daint supercomputer of the Swiss National Supercomputing Centre (CSCS). For comparison, we also ran a 14~M particle simulation, which has a resolution scale of $H/4$ and a numerical $\alpha$ of a few thousandths.  We used length, mass and time units of 1~au, 1~$M_\odot$ and $1/2\pi$ years, respectively. We note that to maintain a balanced computational domain decomposition we also clipped particles with a smoothing length larger than 8~au (equivalent to a density floor of $10^{-16}\,\mathrm{g}\,\mathrm{cm}^{-1}$) in the 14~M particle simulation. We present results at every 430~years ($T$, about $1/3$ of the bending-wave crossing time or 1.2~ORPs).

\section{Results}
\label{sec:results}
\subsection{Disc evolution}

\begin{figure*}
\centering
\includegraphics[width=\linewidth]{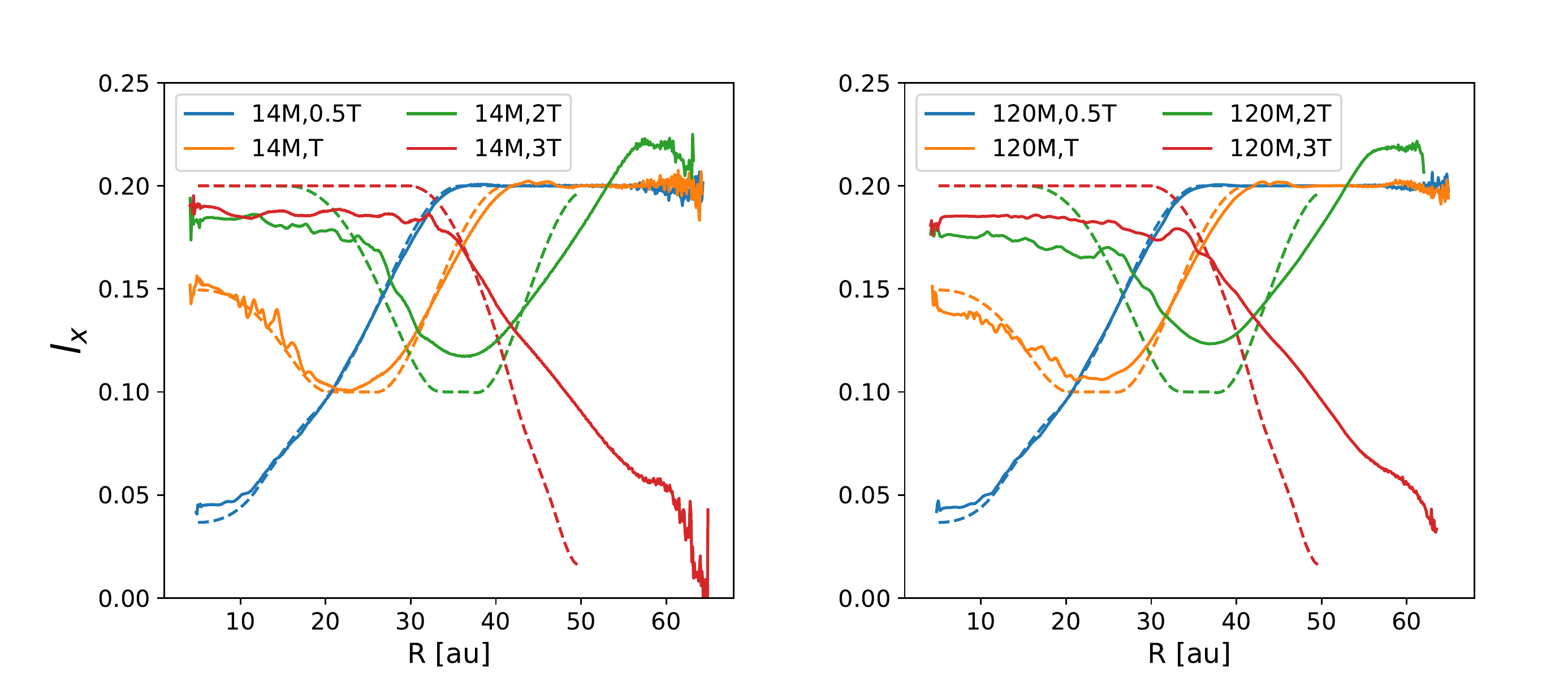}
\caption{The evolution of $l_x$. The dashed lines show predictions from a linear bending-wave solver while the solid lines are the measured $l_x$ within spherical shells. The time unit $T$ equals 430 years. We caution that the linear solution assumes a perfect power-law density profile from 5--50~au with zero torque boundaries while the hydrodynamical simulation has a slightly different density profile (see Fig.~\ref{fig:ic}).  \label{fig:lx}}
\end{figure*}

\begin{figure}
\centering
\includegraphics[width=\linewidth]{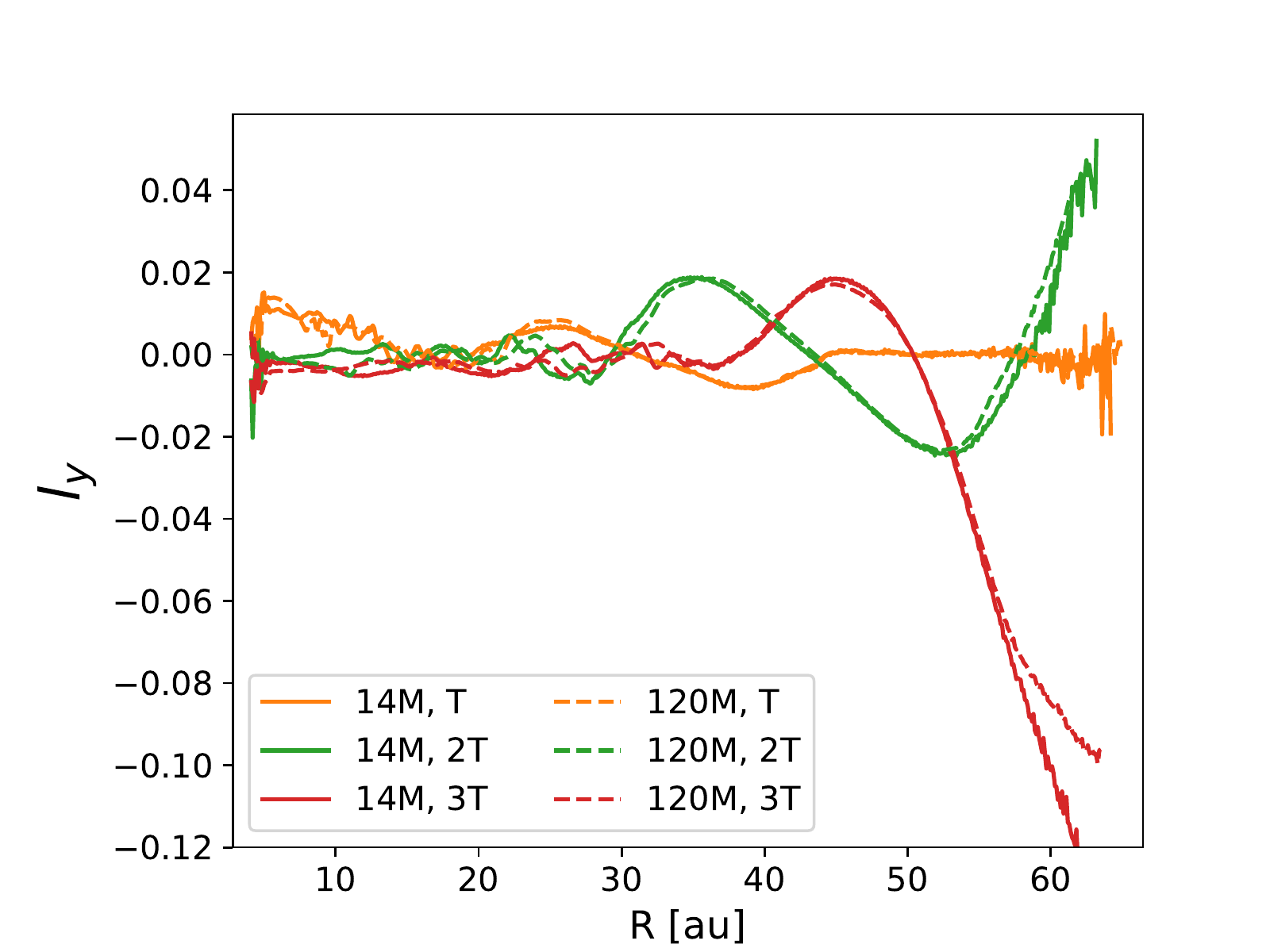}
\caption{The evolution of $l_y$, similar to Fig.~\ref{fig:lx}. The disc is essentially untwisted, with very small $l_y$, before the warp propagates into regions with non-Keplerian rotation beyond 45~au. \label{fig:ly}}
\end{figure}

We plot the gas density in a thin slice near $y=0$ at different times in Fig.~\ref{fig:rho}. The vertical ($z$) axis is parallel to the disc's total angular momentum vector.
The bending wave travels in both directions (see Section~\ref{sec:model} and Appendix~\ref{sec:A}) and produces shocks near the disc surface in the strongly warped region (e.g.\ around 40 au; bottom panel of Fig.~\ref{fig:rho}). We observe weaker shocks in disc surface than \cite{Sorathia2013} due to our smaller warp amplitude $|R\partial{l_x(t)}/\partial{R}| \in (0, 0.6)$.  During the simulation the magnitude of the total angular momentum slightly decreases due to particle clipping ($0.6\%$ and $1.2\%$ by mass for the 14~M and 120~M particle simulations, respectively) at the inner boundary and in the very low-density regions (see Section~\ref{sec:code}).  More particles are clipped in the higher-resolution simulation owing to a more dynamic inner edge. The total angular momentum decreases by 0.8\% and 1.6\% in the low- and high-resolution simulations, respectively, while the specific angular momentum decreases by 0.2\% and 0.4\%. The angular momentum conservation is therefore excellent.

The warp also produces oblique shocks in the body of the disc as seen in the density fields in Fig.~\ref{fig:rho}.  We zoom in to show a spherical shell around 20~au in Fig.~\ref{fig:rholocal}. The high-resolution simulation shows a finer structure, which resembles the density fields in the local simulations of \cite{Gammie2000} and \cite{Paardekooper2019}. These oblique features are characteristic of the inertial waves excited by parametric resonance. However, the low-resolution simulation only vaguely shows some wave activity. The shocks in disc midplane due to the parametric instability are absent in the low-resolution global simulation of \cite{Sorathia2013}.

To follow the propagation of the warp we solve the linear bending-wave equations (see Section~\ref{sec:model}) following \citet{Lubow2002}. We solve these equations in the region 5--50~au with torque-free boundary conditions, assuming a perfect power-law surface density profile. We caution that the solution only provides a reference for the warp evolution before it propagates beyond 50~au. The non-Keplerian rotation beyond 45~au leads to significant precession and twisting (see below). We should also bear in mind the surface density depletion near the inner edge (Fig.~\ref{fig:ic}) in the following interpretation of the results.

We divided the disc into spherical shells of width 0.1 au and then calculated the total angular momentum (both magnitude and direction, i.e.\ tilt vector) of material within each shell. In Fig.~\ref{fig:lx}, we plot the component $l_x$ of the tilt vector at various radii and times and compare this with the linear bending-wave solution. We note that in the tilt vector plot we haven't applied the rotation matrix to align the disc total angular momentum to the $z$ axis as in Fig.~\ref{fig:rho}. The linear bending-wave solution (dashed lines in Fig.~\ref{fig:lx}) agrees perfectly with the measured $l_x$ in the hydrodynamic simulations before the bending wave propagates inwards to 7~au (not shown in Fig.~\ref{fig:lx}). At time $0.5\,T$, the measured $l_x$ in the inner region is larger than the linear solution because the depleted density within 7~au (see Fig.~\ref{fig:ic}) leads to a higher wave amplitude. In general the warp is damped more efficiently in the high-resolution simulation. At time $1~T$, the low-resolution simulation still agrees reasonably well with the linear solution. In the 10--20 au region of the high-resolution simulation, $l_x$ is transferred outwards at a faster pace than the linear solution; the $l_x$ excess around 10~au at $0.5~T$ has been removed by the outward transport of $l_x$. The high-resolution simulation damps the warp faster than the low-resolution simulation owing to more vigorous turbulence ensuing from the parametric instability. We will make that identification in the following section. The difference between the two simulations in the 30--40~au region is more subtle because of the long dynamical time-scale in the outer disk. 
This trend remains at $2~T$ even though the bending wave propagates into the diffusive tail beyond 50~au. We note that in the region beyond 45~au the non-Keplerian rotation is significant and our linear solution is not applicable. The non-Keplerian rotation leads to precession and twisting, which can be seen in the profile of $l_y$ as shown in Fig.~\ref{fig:ly}.

\subsection{Parametric instability}

\begin{figure*}
\centering
\includegraphics[width=0.8\linewidth]{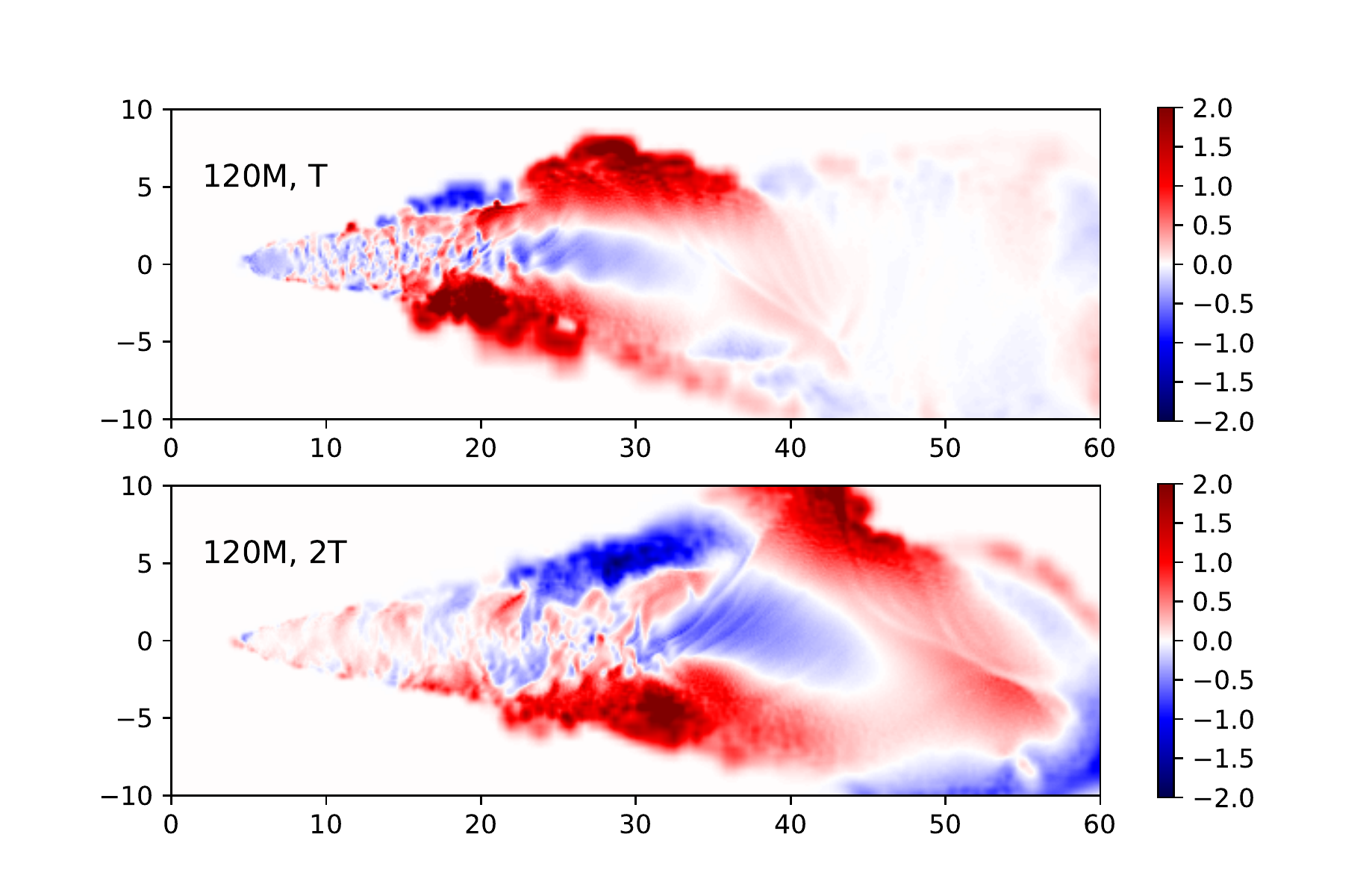}
\caption{Vertical velocity map, normalized to the local sound speed, of the high-resolution simulation shown in Fig.~\ref{fig:rho}. We show only the right panels here for clarity.\label{fig:vz}}
\end{figure*}

Both the density fluctuations and the extra angular momentum transport suggest some instability occurring during the propagation of the bending wave. In this section, we try to pinpoint it. In Fig.~\ref{fig:vz}, we plot the Mach number of the vertical motion (parallel to the $z$ axis) of the gas in the high-resolution simulation of Fig.~\ref{fig:rho}, focusing on the right half of the slice.  The fluid motion is most turbulent in regions with the largest warp amplitude $R|\partial l_x /\partial R|$, i.e.\ around 20~au at $T$ and around 30~au at $2\,T$ (see also Fig.~\ref{fig:lx}). The bending wave also results in shocks in these regions \citep{Gammie2000} as shown by the density field in Fig.~\ref{fig:rho}.

\begin{figure}
\centering
\includegraphics[width=\linewidth]{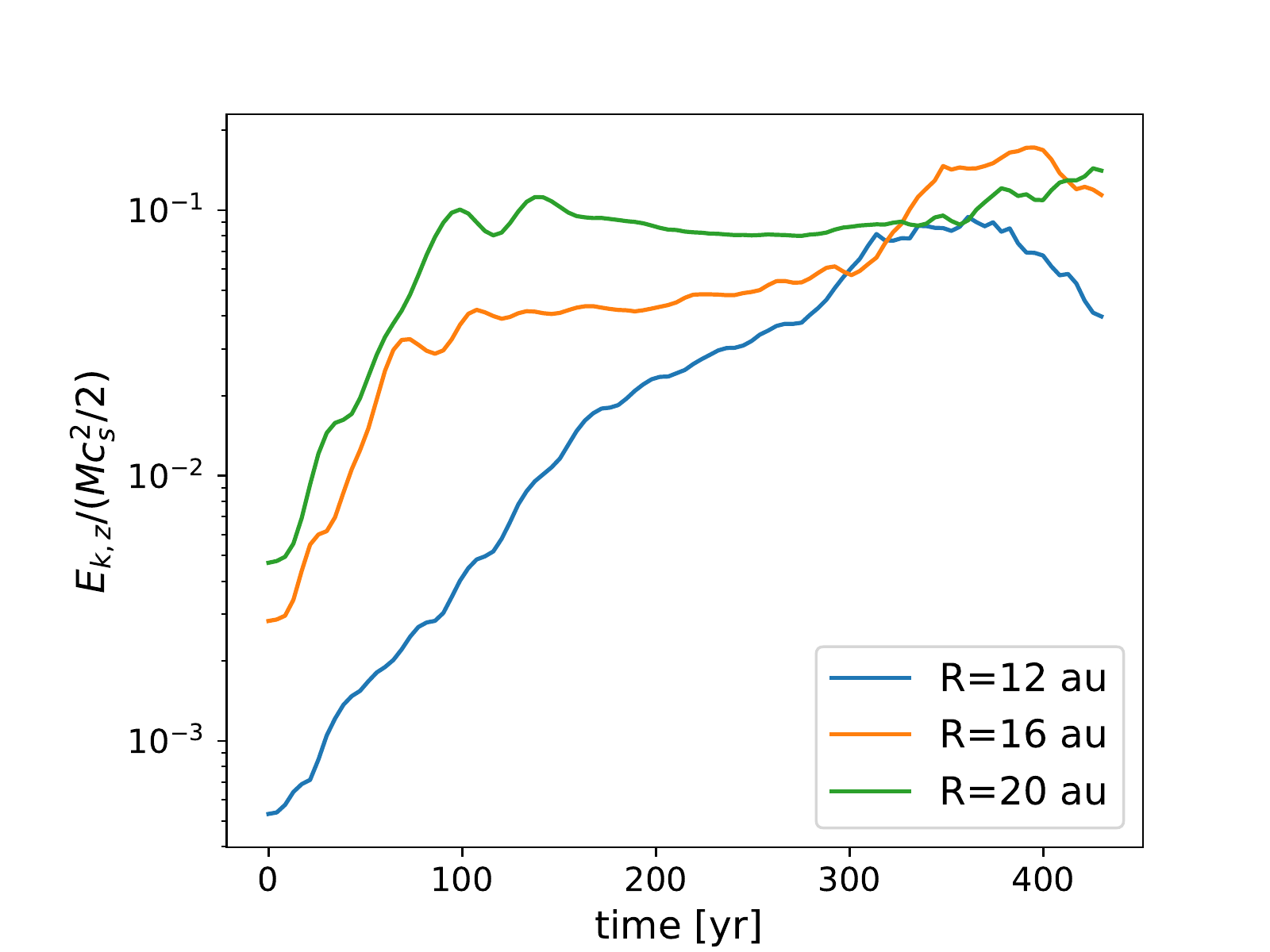}
\caption{Growth of the kinetic energy associated with motion parallel to the local angular momentum vector in thin shells centred at 12, 16 and 20~au.\label{fig:Ezr}}
\end{figure}

The growth of the parametric instability has been studied through local calculations in both the warped shearing box \citep{Ogilvie2013b} and the standard shearing box \citep{Gammie2000}. The growth rate is closely related to the vertical shear rate of the horizontal internal flows in the warped disc. The amplitude of these flows is strongly affected by the resonance between the orbital and epicyclic frequencies in a Keplerian disc. In the warped shearing box model of \citet{Ogilvie2013b}, the amplitude was assumed to be limited by viscous damping and/or non-Keplerian detuning of the resonance, so no prediction was made for the growth rate of the parametric instability in an inviscid Keplerian disc. In a bending wave propagating in an inviscid Keplerian disc, the amplitude of the internal flows is governed instead by time-dependence of the warp, which provides the necessary detuning, according to equations (\ref{bending1})--(\ref{bending2}). The linear analysis of \cite{Gammie2000}
%deals with exact this configuration despite it simplifies the horizontal %flow \citep[see][]{Ogilvie2013b} to an epicyclic motion. %\citep{Gammie2000} found
then shows the growth rate of the parametric instability 
%in inviscid Keplerian flow
to be 
\begin{equation}
s=3\sqrt{3}A/16, \label{eq:gr}
\end{equation}
(the growth rate of energy is $2s$) where $A$ (evolving) is the vertical shear rate as defined in Section~\ref{sec:model}. We note that this is an approximation that is valid only to first order in $A/\Omega$. Their numerical simulation shows a slightly smaller growth rate, either because of insufficient resolution or because  $A/\Omega$ is not $\ll1$ in all cases.

\begin{figure}
\centering
\includegraphics[width=\linewidth]{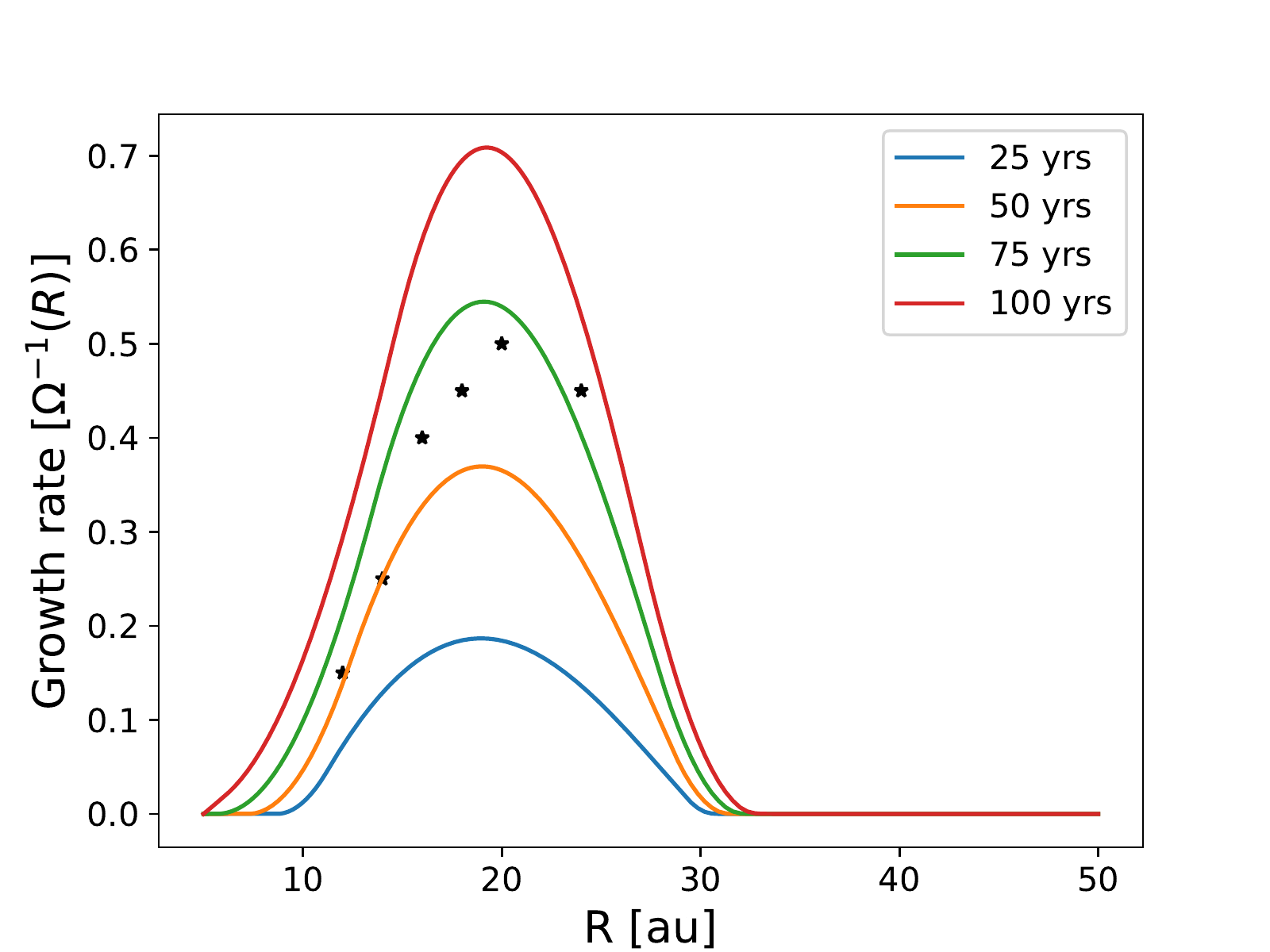}
\caption{The asterisks indicate the fitted mean growth rates of the vertical kinetic energy (t<100 years in figure \ref{fig:Ezr}) at various radii. The curves show predicted instantaneous growth rate at various times based on the shear rate (evolving with time) of the internal horizontal flows (see text). The mean growth rates are in the right range. \label{fig:gr}}
\end{figure}

We carried out a local analysis of the growth of the vertical kinetic energy (see Appendix~\ref{sec:C}) by dividing the disc into spherical shells (of width 1~au) centred at certain radii. We subtracted the mean motion of the shell and aligned its angular momentum to the $z$ axis (see, e.g., Fig.~\ref{fig:rholocal}). We find exponential growth and saturation of the total vertical kinetic energy similar to local shearing-box simulations (see Fig.~\ref{fig:Ezr} and Fig.~\ref{fig:Ekz}). We fit an exponential function to the time series in figure \ref{fig:Ezr} over the first 100 years. The measured growth rates are plotted as asterisks in figure \ref{fig:gr}. In that case we should expect the measured growth rates to be a time-average of the (evolving) theoretical growth rates. We plotted the \emph{theoretical} growth rate (equation \ref{eq:gr}) at various times using  the vertical shear rate in our bending-wave solution in Fig.~\ref{fig:gr} as well. Note we use the local dynamical time-scale, which is radius-dependent, as a unit in Fig.~\ref{fig:gr}. The curves agree qualitatively, especially regarding the peak in the growth rate, indicating that the parametric instability is at work here. 

\subsection{Angular momentum deficit}

\begin{figure}
\centering
\includegraphics[width=\linewidth]{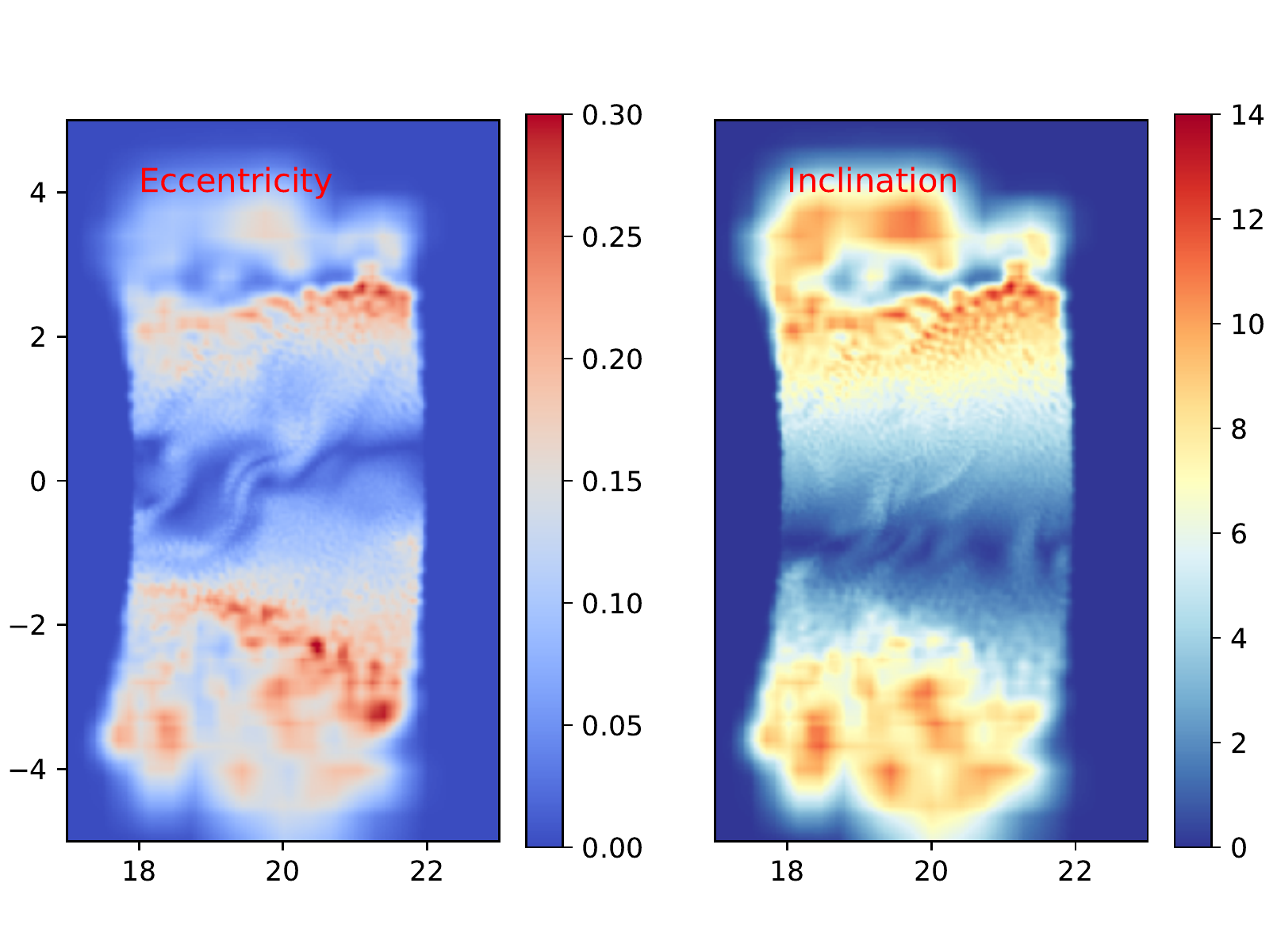}
\caption{Eccentricity and inclination of particles in the high-resolution simulation (zoomed-in plot corresponding to Fig.~\ref{fig:rholocal}). Note that the inclination is shown with respect to the local angular momentum.  \label{fig:ie}}
\end{figure}

The angular momentum deficit (AMD) is widely used in celestial mechanics  \citep{Laskar1997,Laskar2017}. It is a measure of the orbital excitation and is defined, for a system of $n$ particles, as 
\begin{equation}
C=\sum_{k=1}^{n}\Lambda_k\left(1-\sqrt{1-e_k^2}\cos i_k\right),
\end{equation}
where $e_k$ and $i_k$ are the eccentricity and inclination (with respect to the total angular momentum vector) of the $k$~th particle, and $\Lambda_k=m_k(GMa_k)^{1/2}$ is the angular momentum that the $k$~th particle would have if it had a circular orbit with the same semimajor axis $a_k$.
%should it be on a circular orbit aligned with the system's angular %momentum. 
AMD is conserved in the secular theory of celestial mechanics, i.e.\ when mean-motion resonances are unimportant.  Under these circumstances, each particle preserves its orbital energy (and therefore $a_k$ and $\Lambda_k$) but eccentricity and inclination can be exchanged between particles in such a way that $C$ is conserved.

AMD has also been found to be relevant in describing warped and eccentric fluid discs in which inclination and eccentricity can propagate, on timescales much longer than the orbital time-scale, by means of pressure and/or self-gravity.  For example, in the linear theories of \citet[appendix A]{Lubow2001} and \citet[section 3.1]{Teyssandier2016}, we see conservation laws for AMD in systems involving slowly evolving warped or eccentric fluid discs interacting with planets.  In a non-dissipative fluid disc, and in the absence of mean-motion resonances, AMD is conserved because of the conservation of both the orbital energy and the total angular momentum. However, shocks and turbulence can lead to AMD loss. Therefore AMD appears a good indicator for turbulence and shock dissipation. 

In Fig.~\ref{fig:ie}, we plot the local eccentricity and inclination of fluid particles. In a warped disc, the particle orbits can be viewed as a stack of rings that are circular in the midplane but have an eccentricity that increases with altitude. The eccentricity is related to the radial velocity described by the variable $\bm{G}$ in equations (\ref{bending1})--(\ref{bending2}). In a bending wave in a Keplerian disc, both inclination and eccentricity contribute to the AMD, even though the eccentricity vanishes at the midplane; this is seen in equation~A2 of \citet{Lubow2001}. We can also verify directly from equations (\ref{bending1})--(\ref{bending2}) the conservation law for AMD in the linear regime,
\begin{equation}
  \frac{\partial}{\partial t}\left[\frac{1}{2}\Sigma R^2\Omega\left(|\bm{l}|^2+|\bm{e}|^2\right)\right]+\frac{1}{R}\frac{\partial}{\partial R}\left(-\bm{G}\cdot\bm{l}\right)=0,
\end{equation}
where $\bm{l}$ contains the (small) horizontal components of the unit tilt vector, and $|\bm{e}|=2H|\bm{G}|/PR^3=H|A|/R\Omega$ is the (small) eccentricity at one scale-height.

\begin{figure}
\centering
\includegraphics[width=\linewidth]{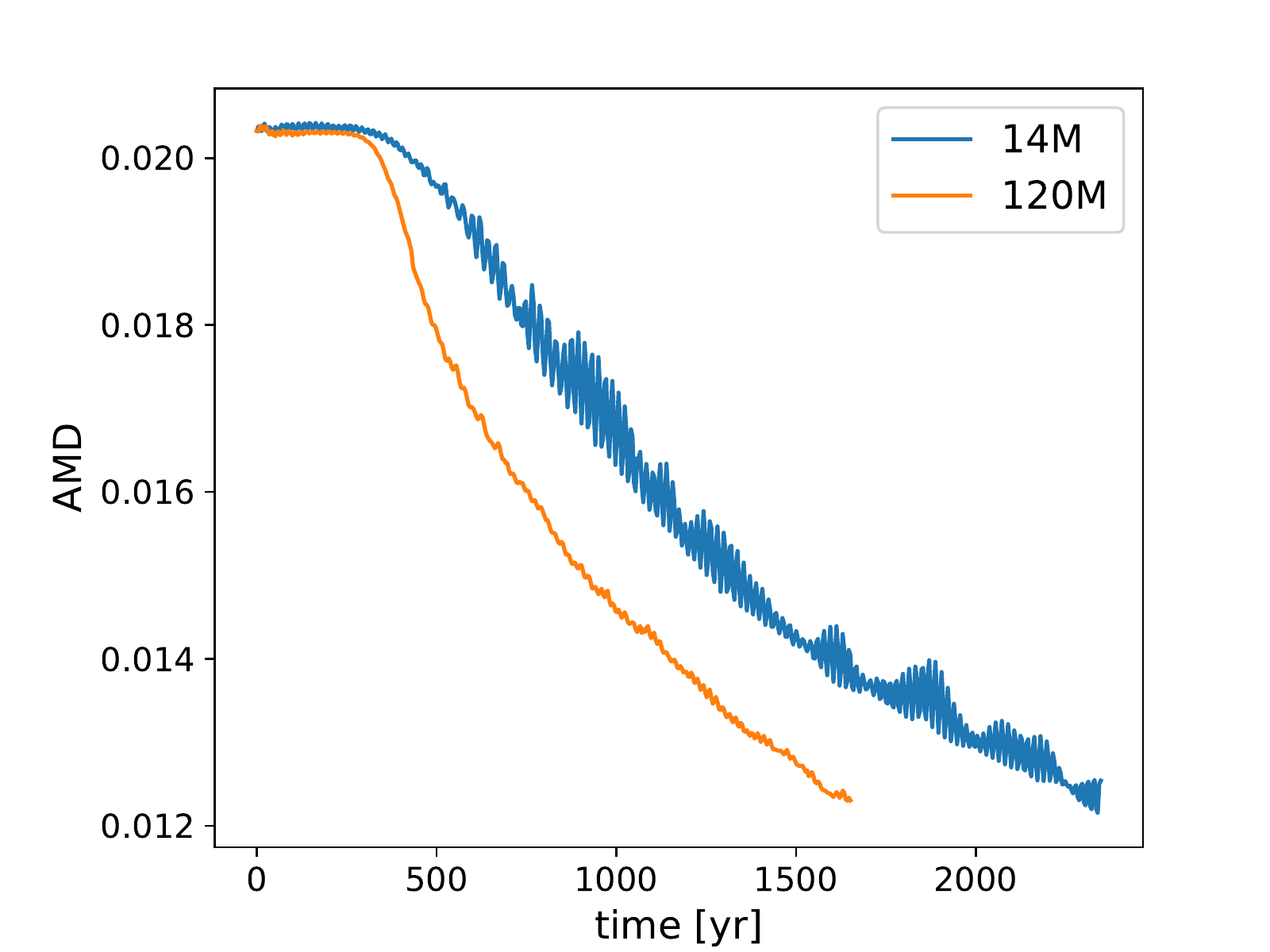}
\caption{The evolution of the averaged AMD, i.e., the total AMD divided by the number of particles. The AMD is conserved before the turbulence saturates and dissipates the orbital energy.  \label{fig:amd}}
\end{figure}

The parametric instability leads to interactions between orbits of different shape and orientation in Fig.~\ref{fig:ie}. In Fig.~\ref{fig:amd}, we plot the AMD evolution in our warped disc simulations. Despite having a lower intrinsic numerical dissipation, the high-resolution simulation shows a faster and stronger AMD decay. During this simulation, the parametric instability leads to shocks and turbulence \citep{Gammie2000} and thus AMD loss (Fig.~\ref{fig:amd}).  The high-resolution simulation shows an earlier and more pronounced AMD decay as a result of a better-resolved parametric instability.

\section{Discussion}
\label{sec:dis}
We find fast bending wave decay due to the parametric instability. With an initial inclination of $11.5^{\circ}$ between the inner and outer parts of the disc, the warp is significantly reduced within one bending-wave crossing time. In a test with an initial inclination of $36.9^{\circ}$,  the AMD decreases by $50\%$ in only one-third of a bending-wave crossing time. This result is in tension with the large group of warped protoplanetary discs indicated by their shadows.

In young circumstellar disc, gravitational instability and magnetohydrodynamic turbulence can generate a relatively large $\alpha$ \citep{Deng2020} leading to diffusive evolution of warps.
 In the diffusive regime with $\alpha>H/R$ one would expect fast warp diffusion on the timescale of  $\alpha(R/H)^2\Omega^{-1}$\citep{Papaloizou1983}. We caution that self-gravity and magnetic fields are likely to affect the warp evolution directly \citep{Papaloizou1995, Paris2018} instead of providing just an effective viscosity, i.e., $\alpha$. In the protoplanetary disc stage, $\alpha$ is then expected to be small in the body of the disc and accretion is likely to be powered by magnetic winds \citep{Bai2013}. We would expect parametric instability to occur in warped discs and to cause a rapid damping of the warp.  Any initial warp is likely to be damped within a few crossing times, so the observed warps are probably maintained by a misalignment in the system, e.g.\ with a companion.

Another possibility is disc breaking. A strongly warped disc may break into independently tilted rings with or without external torques \citep{Lodato2010, Nixon2012, Xiang-Gruess2013, Nealon2016, Zhu2019}. After disc breaking, the inner and outer part mainly communicate angular momentum through gravitational torques with low efficiency.  Indeed some of the observed warped protoplanetary discs are strongly misaligned, suggesting that disc breaking may have occurred \citep[see, e.g.][]{Marino2015,Kraus2020}.

\section{Conclusion}
\label{sec:con}
We have carried out unprecedentedly high-resolution simulations of inviscid warped Keplerian discs to explore hydrodynamic instability caused by the warp. The propagation of warps in the disc is well described by the linear bending-wave equations in the early stages. Parametric instability then develops quickly even in the evolving warp, and the local instability growth rate agrees with previous local studies. The ensuing turbulence leads to rapid angular momentum exchange and damps the warp significantly in one bending-wave crossing time. Angular momentum deficit (a measure of orbital excitation) is damped by turbulent motions and the high-resolution simulation with more vigorous turbulence sees faster damping than the equivalent low-resolution simulation.  Our results suggest that free warps in protoplanetary discs damp rapidly; observed warps are likely to be maintained by companions or are the aftermath of disc breaking. 

\section{Acknowledgement}
H.D. acknowledge support from the Swiss National Science Foundation via an early postdoctoral mobility fellowship. We thank the anonymous referee for comments that helped to improve the clarity of the paper. Simulations were performed on the Piz Daint supercomputer of the Swiss National Supercomputing Centre (CSCS). 

\section{Data availability}
The data underlying this article will be shared on reasonable request to the corresponding author.

\appendix

\section{Analytical solution for the bending wave}
\label{sec:A}

Equations~(\ref{bending1})--(\ref{bending2}) for a small-amplitude bending wave in an inviscid, Keplerian disc can be solved analytically for our disc model with $H/R=\epsilon=0.05$ and $\Sigma H=\text{constant}$ \citep[cf.][]{Ogilvie2006}. Let
\begin{equation}
\tau=\int\frac{2}{c_s}dR=\frac{4}{3\epsilon\Omega} \label{eq:tau}
\end{equation}
be the time for a bending wave (which travels at speed $c_s/2$) to reach radius $R$ from a virtual source at the origin. Then $W$ and $\mathcal{G}$ each satisfy a classical wave equation in the variables $(\tau,t)$ and the general solution is
\begin{align}
&W=f(\tau+t)+g(\tau-t),\\
&\mathcal{G}=\frac{GM\Sigma H}{2}\left[f(\tau+t)-g(\tau-t)\right],
\end{align}
where $f$ and $g$ are functions to be determined, representing inwardly and outwardly propagating bending waves, respectively. If the boundary conditions are that the torque $\mathcal{G}$ vanishes at $\tau=\tau_\text{in}$ and $\tau=\tau_\text{out}$, then the boundaries reflect the bending waves and the relevant solution, corresponding to an initial warp $W_0(\tau)$ at $t=0$ for $\tau_\text{in}\leq\tau\leq\tau_\text{out}$ and no initial torque, and valid up to one bending-wave crossing time, is
%\begin{align}
%&W=\begin{cases}
%  \frac{1}{2}\left[W_0(\tau+t)+W_0(\tau-t)\right],&t\leq\tau-\tau_\text{in},\\
%  \frac{1}{2}\left[W_0(\tau+t)+W_0(2\tau_\text{in}-\tau+t)\right],&t\geq\tau-\tau_\text{in},
%\end{cases},\\
%&\mathcal{G}=\begin{cases}
%  \frac{GM\Sigma H}{4}\left[W_0(\tau+t)-W_0(\tau-t)\right],&t\leq\tau-\tau_\text{in},\\
%  \frac{GM\Sigma H}{4}\left[W_0(\tau+t)-W_0(2\tau_\text{in}-\tau+t)\right],&t\geq\tau-\tau_\text{in}.
%\end{cases}
%\end{align}
\begin{align}
&W=\frac{1}{2}(W_L+W_R),\\
&\mathcal{G}=\frac{GM\Sigma H}{4}(W_L-W_R),
\end{align}
where
\begin{align}
&W_L=\begin{cases}
  W_0(\tau+t),&0\leq t\leq\tau_\text{out}-\tau,\\
  W_0(2\tau_\text{out}-\tau-t),&\tau_\text{out}-\tau\leq t\leq\tau_\text{out}-\tau_\text{in},
\end{cases}\\
&W_R=\begin{cases}
  W_0(\tau-t),&0\leq t\leq\tau-\tau_\text{in},\\
  W_0(2\tau_\text{in}-\tau+t),&\tau-\tau_\text{in}\leq t\leq\tau_\text{out}-\tau_\text{in}.
\end{cases}\\
\end{align}
The long-term solution after multiple reflections at boundaries is less elegant and solved numerically in Section~\ref{sec:model}.

\section{Numerical viscosity}
\label{sec:B}
The bending-wave evolution is sensitive to viscosity, of either physical or numerical origin. We performed shearing-box \citep{Goldreich1965} simulations as described in \citet{Deng2019} to evaluate the level of numerical viscosity at the resolution employed in the global simulations described in the main text. We initialized a bending wave (travelling at half the sound speed) in an isothermal stratified shearing box \citep{Gammie2000} using the initial condition
\begin{align}
    \rho= &\rho_0 \exp\left [-\frac{(z-a \cos(kx))^2}{2H^2}\right ], \\
    v_x= &\frac{a z}{H}\Omega \cos(kx),\\
    v_y=&-\frac{3}{2}\Omega x +\frac{1}{2}\frac{a z}{H}\Omega \sin(kx), \\
    v_z= &a\Omega \sin(kx).
\end{align}
Here $\rho$, $a$, $H$ and $\Omega$ are the density, warp amplitude, scale-height and local angular velocity, respectively. Velocities and coordinates have their usual definitions. We set the wavenumber $k=2\pi/L_x=0.28/H$, where the box with $(L_x, L_y, L_z)=(16\sqrt{2}\,H,2\sqrt{2}\,H,8\,H)$ is resolved by 1 million particles. The midplane particle separation and resolution scale are $0.06\,H$ and $H/8$ using the Wendland C4 kernel \citep{Dehnen2012, Deng2019}. 

\begin{figure}
\centering
\includegraphics[width=\linewidth]{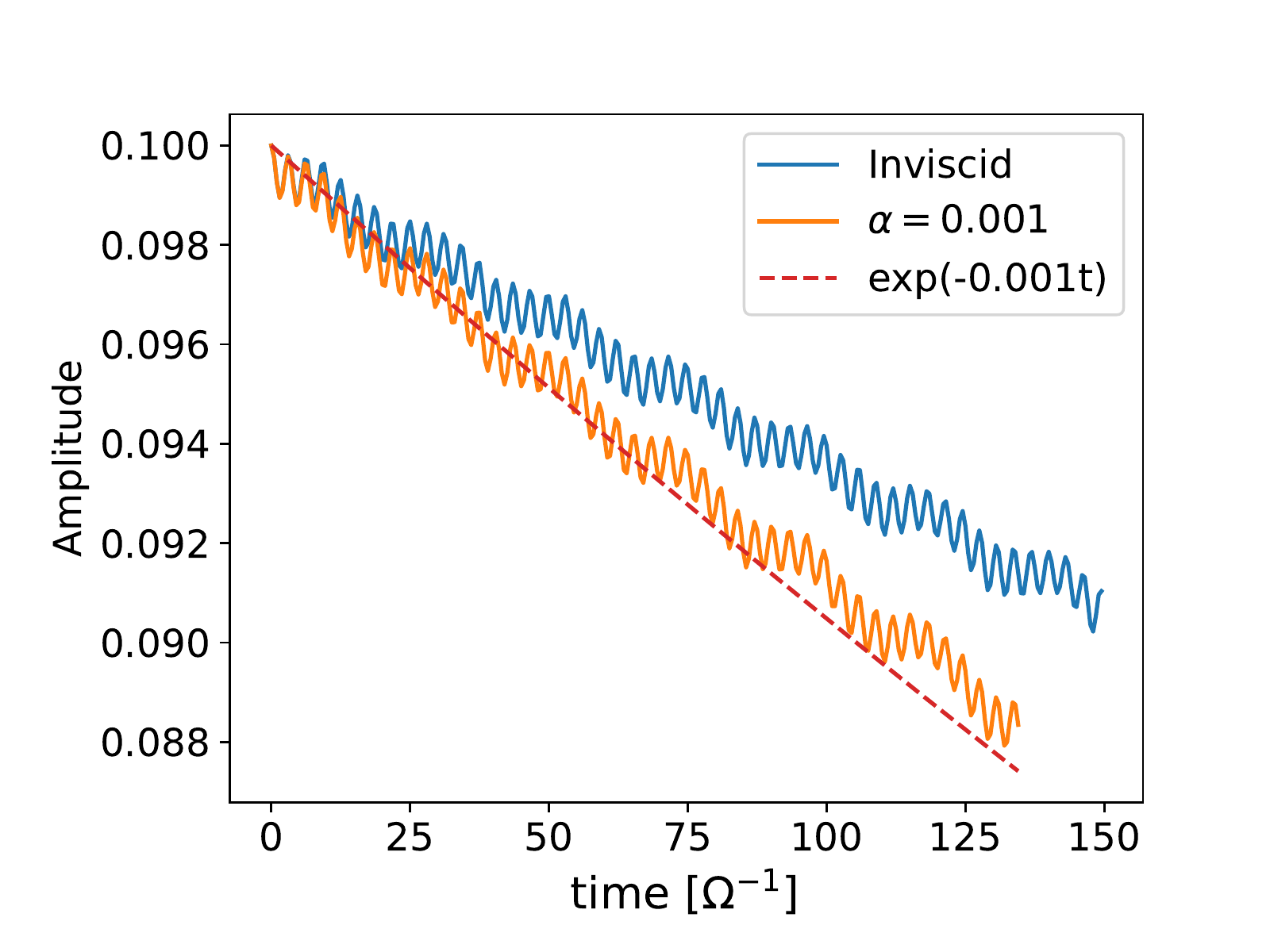}
\caption{Bending-wave decay ($k=2\pi/L_x=0.28/H$ and $a/H=0.1$) with and without explicit viscosity. \label{fig:decay}}
\end{figure}

In the long-wavelength limit, the theoretical bending-wave decay rate is $\alpha \Omega/2$. However, to simulate a very long bending wave with our three-dimensional code would be very expensive. In Fig.~\ref{fig:decay}, we compare the evolution of a bending wave with $k=0.28/H$ and $a/H=0.1$ in an inviscid simulation and a simulation with shear viscosity $\alpha=0.001$ in the full Navier-Stokes equations \citep{Hopkins2017}.  The amplitude of the $m=1$ Fourier mode of the complex warp amplitude, $z+iv_z/\Omega$, is shown in Fig.~\ref{fig:decay}.
The inviscid bending wave decays more slowly than in the $\alpha=0.001$ simulation and we hence conclude that we have numerical viscosity $\alpha<0.001$ at a resolution of $H/8$ in the midplane of the vertically stratified simulations. We note that a similar simulation with half the linear resolution ($1/8$ the number of particles) decays slightly faster than the $\alpha=0.001$ curve. 

\section{Local simulation of parametric instability}
\label{sec:C}
\begin{figure}
\centering
\includegraphics[width=\linewidth]{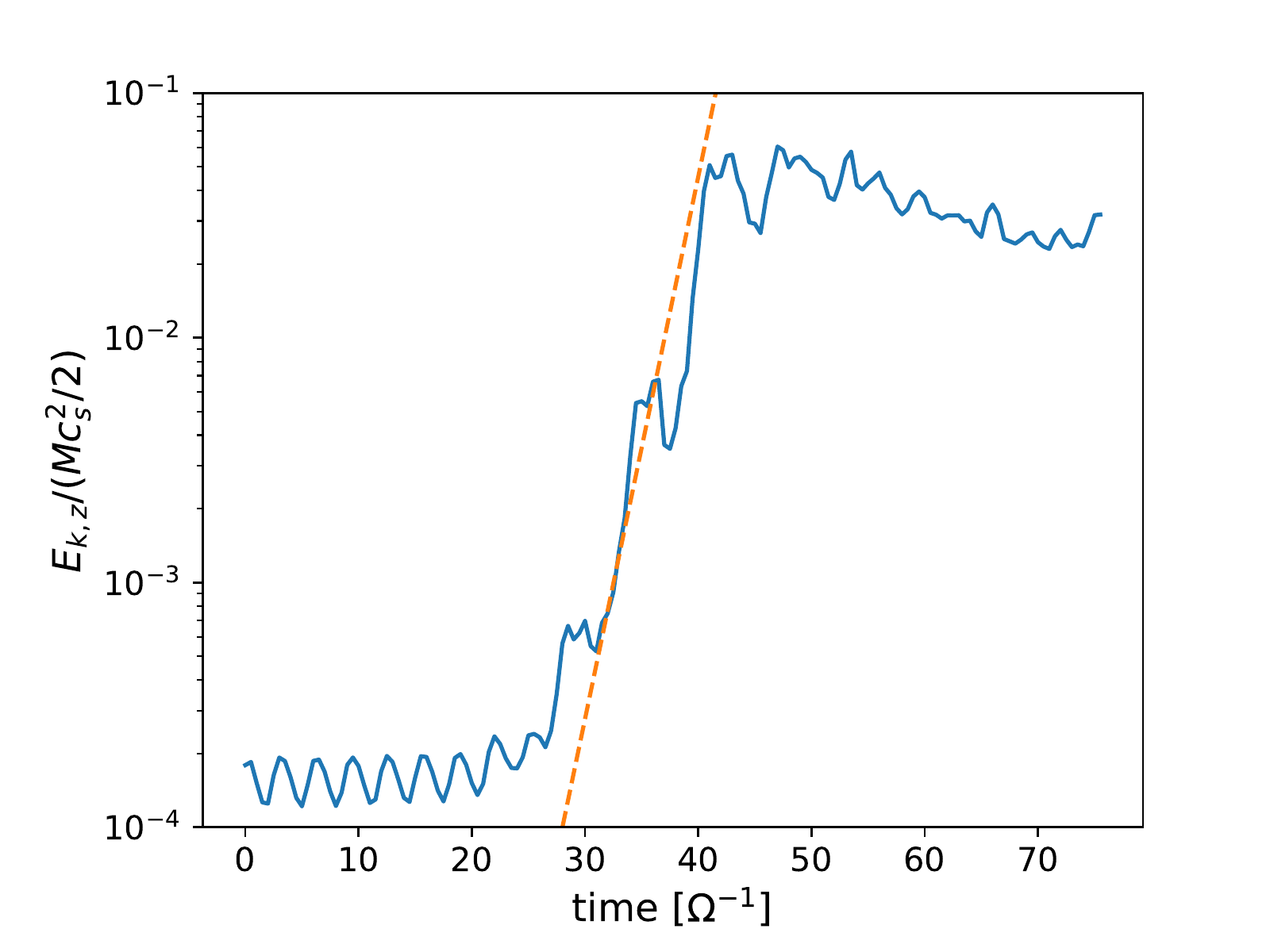}
\caption{Solid line shows the growth of the vertical kinetic energy as an indicator of the parametric instability. Dashed line indicates an exponential growth with a growth rate of $0.51\Omega$.\label{fig:Ekz}}
\end{figure}

Our Lagrangian method has an adaptive resolution so that the denser regions are sampled with more particles and thus better resolved. A dedicated resolution study is needed instead of using grid-code resolution criteria. We ran the fiducial local simulation of parametric instability in \cite{Gammie2000}. We sample a box of dimension $(L_x, L_y, L_x)=(4\,H,16\,H,8\,H)$ with 1 million particles. The midplane resolution scale is again $H/8$. A radial velocity $v_x=\Omega z$ is initialised and the parametric instability develops quickly. 
In Fig.~\ref{fig:Ekz}, we plot the evolution of vertical kinetic energy normalised to half the internal energy. Our Lagrangian code has larger initial noise in the velocity field than the grid simulations of \cite{Gammie2000}.  However, we measured a numerical parametric instability growth rate of $0.51\Omega$ as in the fiducial model of  \cite{Gammie2000} (who used a finite difference method at a resolution of $H/32$) at a disc midplane resolution of only $H/8$. We note that the numerical growth rate is smaller than the theoretical prediction (see equation \ref{eq:gr} and related comments).

\bibliographystyle{mnras}
\bibliography{references}

%%%%%%%%%%%%%%%%%%%%%%%%%%%%%%%%%%%%%%%%%%%%%%%%%%

%%%%%%%%%%%%%%%%% APPENDICES %%%%%%%%%%%%%%%%%%%%%
% Don't change these lines
\bsp	% typesetting comment
\label{lastpage}
\end{document}